# Evolutionary Circuit Design: Information Theory Perspective on Signal Propagation

*Denis Popel*
Department of Computer Science,
Baker University,
P.O. 65, Baldwin City, KS 66006,
E-mail: `popel@ieee.org`

*Nawar Hakeem*
Department of Computer Science,
University of Wollongong (Australia),
Dubai Campus, P.O. Box 20183, Dubai, U.A.E.,
E-mail: `hakeem@ieee.org`

*Abstract*— This paper presents case-study results on the application of information theoretic approach to gate-level evolutionary circuit design. We introduce information measures to provide better estimates of synthesis criteria of digital circuits. For example, the analysis of signal propagation during evolving gate-level synthesis can be improved by using information theoretic measures that will make it possible to find the most effective geometry and therefore predict the cost of the final design solution. The problem is considered from the information engine point of view. That is, the process of evolutionary gate-level circuit design is presented via such measures as *entropy*, *logical work* and *information vitality*. Some examples of geometry driven synthesis are provided to prove the above idea.

## I. Introduction

The impact of different design strategies on the time cycle of digital circuit creation, and the effect of performance and power dissipation on the final realization have increased the importance to integrate the technology - independent and technology - dependent design stages. In other words, logic specification should become closer to the final topological structures and physical implementation of digital circuits. To move in this direction, we address the problem of automatic synthesis implementations using gate-level evolutionary approach [2]. Gate-level evolutionary design is of great interest to researchers as it outlines the major steps to synthesize the final realization [3]. Concepts and techniques developed for different design strategies separately can therefore be exploited simultaneously in evolutionary circuit design. Our study is based on the pioneering work of the gate-level evolutionary synthesis which exploits geometry representation of rectangular array of cells [2], [4].

In information networks, of which digital circuits are an integral component, one deals with a variety of logical operations and complex processes delivering signals before information reaches the final destination. These information systems seem to need a new basic theory which helps to understand the system in its entirety in order to provide a basis for designing the total system. Such a theory was developed in [8]. In accordance with the theory, the machines which deal with information can be generally referred to as *information engines* which model the actual information systems. In this paper, we propose such an interpretation for evolutionary circuit design which considers stochastic behaviour of evolutionary synthesis in the context of information measures [1], [7]. Hence, entropy is used as a complexity characteristic in the circuit-design optimization process and generally in signal processing [5]. Altogether, we describe a new method of analysis for geometry-based evolutionary synthesis in which the main contribution is using information theoretic measures for calculating parameters of geometry in evolutionary gate-level circuit design.

This paper is structured as follows. Section II explains the background. Section III describes the information theory interpretation of geometry-based gate-level circuit design. We introduce information measures and their applications to digital circuit design in Section IV and show case-study results. We conclude and open a discussion forum in Section V.

## II. Preliminaries and Assumptions

Consider logic representation of a digital circuit in the form of *Boolean function f* treated as the mapping $\{0,1\}^n \rightarrow \{0,1\}^m$ over the variable set $X = \{x_1, \cdots, x_n\}$. Here, $n$ is the number of variables (inputs), and $m$ is the number of functions (outputs).

## A. Evolutionary Circuit Design

The central idea of evolutionary gate-level circuit design is to search for functionality and connectivity of an array of cells whose parameters are defined by the circuit layout called *circuit geometry* [4]. Possible solutions are represented using chromosomes which are exploited by genetic algorithm with standard operations such as mutation, crossover and selection. The length of a chromosome depends on circuit geometry. In our work, the circuit geometry can be changed, therefore a chromosome of variable length is considered.

The evolution of the digital circuit design is twofold: first, 100% functionality has to be obtained; second, the complexity of realization, i.e., number of gates, inter-connections and layout should be taken into consideration. It was shown in [3] that by allowing the circuit geometry to be flexible, we can evolve circuits with the smallest number of used gates. Other circuit geometry characteristics are: number of inputs, size of the array of logic cells in the form of the number of rows and columns, degree of connectivity defined by a level-back parameter, and the functionality or the fitness function.

In this paper, we consider only two aspects of evolutionary circuit synthesis: determining the size of $n \times m$ cellular array with $n_I$ inputs and $n_O$ outputs, and defining the library of gates needed for the process of evolutionary circuit design. The circuit layout is given as a rectangular array of gates which are uncommitted (Figure 2(a)). We restrict our investigation to the cell library $\alpha$ of basic logic functions $\mathcal{L} = \{\mathcal{NOT}, \mathcal{AND}, \mathcal{OR}, \mathcal{EXOR}\}$ with primary and inverted inputs. We omit the stages of evolutionary synthesis and further verification of evolved circuit. The efficiency of the search process is indicated using the so called fitness function. The fitness function which employs information theoretical measures was proposed in [3]. In our study of geometry, we utilize information theoretic notations as a tool to quantify the characteristics of geometry and cell library.

## B. Information Theoretic Notations

In order to quantify the content of information for a finite field of events $A = \{a_1, a_2, \cdots, a_n\}$ with probabilities distribution $\{p(a_i)\}$, $i = 1, 2, \cdots, n$, Shannon introduced the concept of *entropy* [7]: $H(A) = -\sum_{i=1}^{n} p(a_i) \cdot \log p(a_i)$, where log denotes the base 2 logarithm. For two finite fields of events $A$ and $B$ with probability distribution $\{p(a_i)\}, i = 1, 2, \cdots, n,$ and $\{p(b_j)\}$, $j = 1, 2, \cdots, m$, probability of the joint occurrence of $a_i$ and $b_j$ is joint probability $p(a_i, b_j)$, and there is conditional probability, $p(a_i|b_j) = p(a_i, b_j)/p(b_j)$. The *conditional entropy* of $A$ given $B$ is defined by $H(A|B) = -\sum_{i=1}^{n} \sum_{j=1}^{m} p(a_i, b_j) \cdot \log p(a_i|b_j)$.

In case of logic networks and signal processing, we assume that the sets of values of a function $f$ and arbitrary variable $x$ are two finite fields [5]. We calculate the probability $p_{|f=b} = k_{|f=b}/k$, where $k_{|f=b}$ is the number of assignments of values to variables for which $f = b$ and $k$ is the total number of assignments.

*Example 1:* For the function $f = \overline{x}_3 \cdot \overline{x}_2 \vee x_1$ with truth vector [10001111]: $H(f) = -5/8 \cdot \log_2(5/8) - 3/8 \cdot \log_2(3/8) = 0.96$ *bit*, $H(f|x_1) = -1/8 \cdot \log_2(1/4) - 3/8 \cdot \log_2(3/4) - 4/8 \cdot \log_2(4/4) - 0 = 0.41$ *bit*. By the same computations $H(f|x_2) = 0.91$ *bit*, $H(f|x_3) = 0.91$ *bit*.

## III. INFORMATION ENGINE OF CIRCUIT EVOLUTION

Consider an information engine with one input port and one output port, where the initial description of the logic function $f$ is converted to output description, *evolvable circuit*. Any process done by the information engine can be considered as a composition of two types of processes, *isentropic* process and entropy changing process. Successful (100% functionality) evolutionary design is an isentropic process that does not change the function itself. When we deal with an entropy changing process that requires the conversion of input and output combinations or ensembles, we call such a process *iso-vitality*.

The concept of information engine of evolvable gate-level circuit design is based on the following notation. For every input description of the logic function $f$, there exists the logical

*NetWork* (*NW*) which converts input ensemble to output ensemble. Hence, the loss of uncertainty or network information can be characterized by $I_{NW} = H_{loss}(f|X) = H(X) - H(f)$. For each input $x_i$ of the logical network, the value of information transmitted to the output $f$ is $H_{trans}(x_i) = H(f|x_i)$.

### A. Evolutionary Design as Isentropic Process

Consider a circuit design process as a sequence of steps which operate with input variable $X$, and correspond to network representation of the function $f$. Let $q(f; NW)$ be the *logical work* required to convert initial description of $f$ to the network $NW$ via an information engine. The concept of logical work for digital networks was also discussed in [1]. The circuit design process is reversible, i.e., it is possible to produce the initial function description from the network, which means it is an isentropic process: $H(f) = I(f; NW)$ and $I(X; NW) = H(X)$.

For an isentropic process of circuit design, we can use the information potential measure defined next.

*Definition 1:* Information Potential is the inferior bound of quantity of logical work $q(f; NW)$ under all possible $NW$ design processes,
$$Q(f; NW) = Inf[q(f; NW)] \ for \ all \ possible \ NW$$
This is the information potential of logic function $f$ with respect to network $NW$.
Another important information concept is,

*Definition 2:* Information Vitality is the information potential of an ensemble $X$ with respect to network $NW$ per unit of information,
$$T(NW) = Q(f; NW)/H(f)$$
This is the information vitality equation.
The information potential characterizes the cost of final realization of the logic function $f$, therefore, the search of the inferior bound of $q(f; NW)$ represents a search of the logic function that will minimize the cost of realization.

The dynamic behaviour of an information engine of gate-level circuit design can be best described in the entropy vs. vitality diagram, so called H-T-Diagram (Figure 1). The design of digital circuit is an isentropic process with respect to inputs and outputs. Information verification of circuit design is an iso-vitality pro-

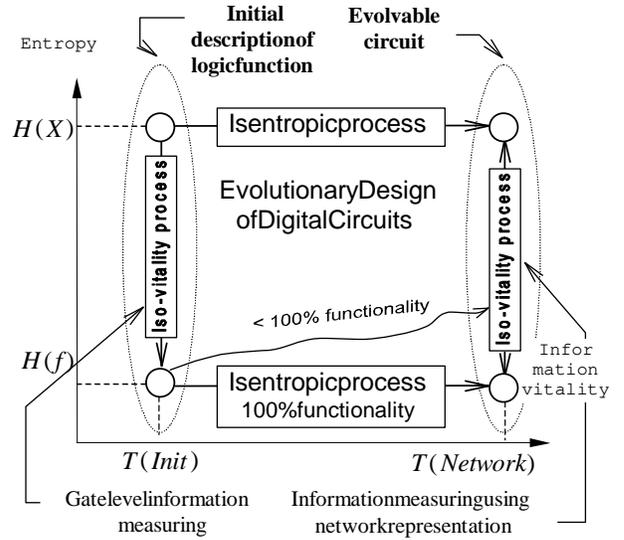

Fig. 1
H-T-Diagram of gate-level circuit design process cess. We can evaluate the loss of information via construction of a logic network for the verified function. The synthesis of an information engine for evolutionary gate-level circuit design deals with all parts of H-T-Diagram in addition to information measuring which is an essential part of the synthesis process.

### B. Gate-level Information Measuring

Let us consider simple gates and its information content in order to analyze the information transmission through each gate. The concept of *logical work* is crucial for information measures of Boolean networks [1]. We use the following notation for each gate: the input entropy $H(X)$, and the output entropy of function $f$ is $H(f)$.

*Definition 3:* The conditional entropy $H(f|x)$ between arbitrary variable $x$ and the function $f$ is a measure of transmission of information from input $x$ to the output $f$.

*Definition 4:* The difference between input and output entropies of a gate,
$$I_{gate} = H(X) - H(f),$$
is an information theoretic measure of the *gate*.

*Example 2:* For the function $\mathcal{AND}$ we obtain the following entropy measures: $H(X) = -^1/_4 \cdot \log{^1/_4} - ^1/_4 \cdot \log{^1/_4} - ^1/_4 \cdot \log{^1/_4} - ^1/_4 \cdot \log{^1/_4} = 2$ bits, $H(f) = {}^1/_4 \cdot \log {}^1/_4 - {}^3/_4 \cdot \log {}^3/_4 = 0.81$ bits, $I_{gate} = 2 - 0.81 = 1.19$ bits, $H_{trans}(f|x_1) = H_{trans}(f|x_2) = 0.5$ bits.

Note, the uncertainty of the output of a digital network is never increased by knowledge of input patterns. In other words, the digital network does not increase information carried by the inputs.

## IV. Circuit Layout Measuring

We focus in our study on a scanning process for a specified geometry style to evolve digital circuits defined over a *library* $\mathcal{L}$ of gates $\mathcal{G}_1$, ..., $\mathcal{G}_t$, where $t$ is a total number of gates in the library. We consider geometry $p \times q$ as one of the main components of the evolutionary design specification. Note that geometry incorporates many parameters: the maximal number of $p$ levels, the maximal number $q$ of gates in each level of the evolved network, types of logic gates from the standard logic library $\mathcal{L}$, network architecture (permissible interconnections between cells, levels, inputs and outputs of the circuit).

In the study of gate-level evolutionary design, the desired logic function has to be implemented through the scanning process, i.e. processing of the full circuit is accomplished by dividing it into subparts (windows) which are scanned in turn, to fulfill given restrictions on library gates and geometry structure. The runtime and the final cost of circuit realization should be minimal. It is essential to consider different geometry realizations, and various schemes of assigning logic elements to different circuit levels (see, for example, $\mathcal{AND} - \mathcal{OR} - \mathcal{EXOR}$ networks in [6]).

We will vary the number of levels, gates in each level, and types of gates, and assign gates to many circuit levels in order to optimize the evolution process. We investigate the fact shown in [4] that evolving some circuits is easier to do on a larger scanning window, but this has resulted in a less efficient circuit.

The evolution of an $q$-level, $n$-input, $m$-output circuit, over the fixed library of cells $\mathcal{L} = \{\mathcal{G}_1, \ldots, \mathcal{G}_t\}$ is defined by the scanning window $p \times q$. The structure of $q$-level circuit is characterized by the number of levels allowed to be connected with inputs and outputs, such parameter is called *level back*. So, the result of this searching process is a set of circuits with no more than $q$ levels and no more than $p$ gates in each level.

We estimate a scanning window size based on the information content with the final goal to optimize this window, or at least to give useful recommendations on how to choose the geometry to design a circuit in accordance with the given specification.

### A. Information Measures of Library Cells

The main idea of information measuring of a geometry is to give numerical estimations and establish propositions for the scanning parameters. Here we define *information capacity* for a gate, a library of gates, and geometry.

We select gates, for every position of the scanned window, from the cell library $\mathcal{L}$. The information equivalents of library gates are given in Table I. We consider $I_{gate}$ as an *information capacity* of the gate (Figure 2). Here, we do not take into account the information carried by inputs and outputs of the network, because such an information does not influence geometry parameters.

*Definition 5:* Information capacity $I_\mathcal{L}$ of a library $\mathcal{L}$ of gates $\mathcal{G}_1, \ldots, \mathcal{G}_t$ is the entire amount of information capacities of the gates:

$$I_\mathcal{L} = \sum_\mathcal{L} I_{gate}.$$

*Example 3:* Given the gate library $\mathcal{L} = \{\mathcal{NOT}, \mathcal{AND}, \mathcal{OR}\}$, information capacity is $I_\mathcal{L} = I_{gate}(\mathcal{NOT}) + I_{gate}(\mathcal{AND}) + I_{gate}(\mathcal{OR}) = 2.38$ bit.

*Definition 6:* Information capacity $I_\mathcal{G}$ of a single cell with respect to a scanning window is the maximal information capacity of all gates from the given (fixed) cell library $\mathcal{L}$:

$I_\mathcal{G} = max\{I_{gate}\}$ for all gates from library $\mathcal{L}$.

Based on these definitions, we can assume that information capacity of $p \times q$ geometry over a fixed library $\mathcal{L}$ of gates $\mathcal{G}_1, \ldots, \mathcal{G}_t$ is a composition of information capacities of the cells:

$$I_{Geometry} = p \cdot q \cdot I_\mathcal{G}$$

The information capacity of a geometry also depends on interconnections among the cells, and the level back parameter with respect to inputs and outputs of an evolving circuit. We use rough estimations of information capacities for each level of the created circuit based on information measures. Then, proceed from this as-sumption and additional factors, in particular, the transmitted information through each level of the created circuit in order to estimate information capacity of the geometry of a given logic function, and analyze the gate library suitability in terms of design parameters in our evolutionary algorithm.

TABLE I
INFORMATION EQUIVALENT OF PRIMITIVE GATES IN BITS

| Library gates | $\mathcal{NOT}$ | $\mathcal{AND}$ | $\mathcal{OR}$ | $\mathcal{EXOR}$ |
|---|---|---|---|---|
| | | | | |
| Function $f$ | [01] | [0001] | [0111] | [0110] |
| Maximum input information $H(X)$ | 1.0 | 2.0 | 2.0 | 2.0 |
| Output information $H(f)$ | 1.0 | 0.81 | 0.81 | 1.0 |
| Gate information measure $H(X)-H(f)$ | 0.0 | 1.19 | 1.19 | 1.0 |
| Transmission of information $H(f|x)$ | 1.0 | 0.5 | 0.5 | 1.0 |

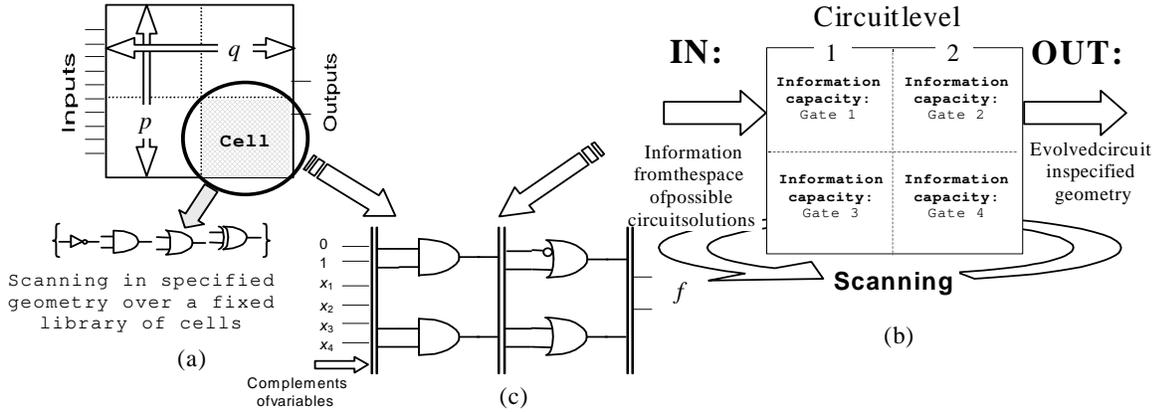

Fig. 2
EVOLUTIONARY GATE-LEVEL CIRCUIT SYNTHESIS REALIZED BY SCANNING USING GEOMETRY AND GIVEN LIBRARY OF GATES

*Example 4:* Given the cell library $\mathcal{L} = \{\mathcal{NOT}, \mathcal{AND}, \mathcal{OR}\}$ and a scanning window of $2 \times 2$, the information capacity of cells in the first level is $I_\mathcal{G} = 1.19$ $bit$, and in the second level is $I_\mathcal{G} = 0.595$ $bit$. The information capacity of the geometry is equal to $I_{Geometry} = 3.57$ $bit$ (Figure 2(c), see corresponding geometry in Figure 3(c)).

### B. Proper Selection of Geometry: Case Study

We summarize the information measures for different specifications in Table II. These mea-sures allow us to make a priori decision about the efficiency of a scanning window.

TABLE II
INFORMATION CAPACITY OF GEOMETRIES

| | Geometry | | |
|---|---|---|---|
| | $\{\mathcal{NOT},$ $\mathcal{AND},\mathcal{OR}\}$ | $\{\mathcal{NOT},$ $\mathcal{EXOR}\}$ | $\{\mathcal{NOT},\mathcal{AND},$ $\mathcal{OR},\mathcal{EXOR}\}$ |
| | $2 \times 2$ | | |
| $I_{Geometry}$ | 3.57 | 3.00 | 3.57 |
| | $3 \times 3$ | | |
| $I_{Geometry}$ | 6.2475 | 5.2500 | 6.2475 |

*Example 5:* Let us estimate the scanning process for an evolving 4-input, 2-output digital circuit. For the purpose of simplification, assume these specifications,

*Specification 1:* $3 \times 3$ geometry over the gate library $\mathcal{L} = \{\mathcal{NOT}, \mathcal{EXOR}\}$. The informa-tion capacity of the given geometry a priori is $I_{Geometry} = 3 \cdot 1.00 \cdot (1 + ^1/_2 + ^1/_4) = 5.25\ bit$ (Figure 3(a)). Let us assume that it is possible to realize the circuit with a maximum of 6-inputs and 3-outputs, but during the design 4-input, 2-output circuit was created. It means that some of cells $3 \times 3$ did not get utilized (Figure 3(b)). For this case, the information capacity of the geometry is $I_{Geometry} = 3.5\ bit$.

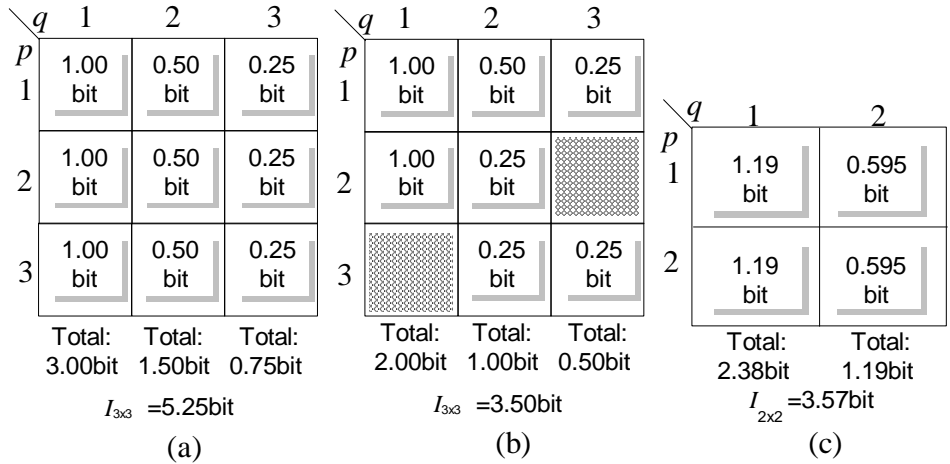

Fig. 3
ESTIMATIONS OF GIVEN GEOMETRY FOR EXAMPLE 5

*Specification 2:* $2 \times 2$ geometry over the gate library $\mathcal{L} = \{\mathcal{NOT}, \mathcal{AND}, \mathcal{OR}\}$. The information capacity of the design style is $I_{Geometry} = 2 \cdot 1.19 \cdot (1 + ^1/_2) = 3.57\ bits$. Such geometry allows to evolve 4-input, 2-output circuits (Figure 3(c)).

Hence, $2 \times 2$ geometry over $\mathcal{L} = \{\mathcal{NOT}, \mathcal{AND}, \mathcal{OR}\}$ has greater information capacity than $3 \times 3$ geometry over $\mathcal{L} = \{\mathcal{NOT}, \mathcal{EXOR}\}$ to evolve 4-input, 2-output circuit.

## V. CONCLUDING REMARKS

It has been shown that information theory measures of the evolutionary circuit design process are useful and give new possibilities to improve the efficiency of the recently developed evolutionary techniques. The main contribution of this paper is the inclusion of circuit information content in the evolutionary scanning process and signal propagation in the context of an information engine. The extension of the recently developed technique of evolutionary circuit design includes information quantification of a cell library, and estimation of circuit geometry. Hence, results obtained with the proposed technique show that it can be useful for a priori analysis of gate-level evolutionary synthesis.

## REFERENCES

[1] L. Hellerman. A measure of computation work. *IEEE Trans. Computers*, C-21(5):439–446, 1972.
[2] H. Iba, M. Iwata, and T. Higuchi. Machine learning approach to gate level evolvable hardware. In *Lecture Notes in Computer Science, Springer-Verlag, vol. 1259*, pages 327–393, 1997.
[3] T. Luba, C. Moraga, S. Yanushkevich, M. Opoka, and V. Shmerko. Evolutionary multi-level network synthesis in given design style. In *Proc. of the IEEE Int. Symp. on Multiple Valued Logic*, pages 253–258, 2000.
[4] J. F. Miller, P. Thomson, and T. Fogerty. *Designing Electronic Circuits Using Evolutionary Algorithms. Arithmetic Circuits: A Case Study*, pages 105 – 131. John Wiley and Sons Ltd, 1998.
[5] D. Popel. Towards efficient calculation of information measures for reordering of binary decision diagrams. In *Proc. of the IEEE Int. Symp. on Signals, Circuits and Systems*, pages 509–512, 2001.
[6] T. Sasao. *Switching Theory for Logic Synthesis*. Kluwer Academic Publishers, 1999.
[7] C. Shannon. A mathematical theory of communication. *Bell Syst. Tech. J.*, 27:379–423, 623–656, 1948.
[8] H. Watanabe. Network theory of information engine. In *Proc. European Conf. on Circuit Theory and Design - ECCTD'97*, pages 6–11, 1997.